# PHOTOEFFECT IN POLYTHIOPENTACENE FILMS AND INFLUENCE OF PERMANENT ILLUMINATION ON IT

M.P. GORISHNYI

UDC 535.343.2;535.343.3;537.311.322;537.312.5.
Institute of Physics, Nat. Acad. Sci. of Ukraine
(46, Nauky Prosp., Kyiv 03680, Ukraine; e-mail: gorishny@iop.kiev.ua)

The internal photoeffect in polythiopentacene (PTP) films is investigated. It is discovered that, in capacitor cells, the spectral dependence of the frontal photo-emf $V_f$ correlates with the PTP absorption spectrum, while that of the back photo-emf $V_b$ anticorrelates with it. The polarities of $V_f$ and $V_b$ coincide. In this case, the charge of the free PTP surface is negative, whereas that of the frontal $SnO_2$-electrode is positive. $V_f = V_{Df} - V_{df}$ and $V_b = V_{db} - V_{Db}$, where $V_{Df}$ and $V_{Db}$ are diffusion components, $V_{df}$ $V_{db}$ are drift frontal and back components, respectively. The quantities $V_{df}$ and $V_{db}$ are caused by the antiblocking bending of the bands for holes near the free PTP surface and the back $SnO_2$-electrode, respectively. The strong-absorption monochromatic frontal illuminations with the energies of quanta of 1.73 and 2.67 eV increase the absolute value of $V_{df}$, which decreases the resulting $V_f$ in the spectral range 1.4–2.2 eV and changes its polarity in the range 1.4–1.6 eV. The low-absorption frontal illuminations with the energy of quanta of 1.54 and 2.15 eV decrease $V_{df}$ in the spectral range 1.6–2.0 eV and increase it in the range 1.4–1.6 eV. If the intensity of these illuminations $I_{pi} \geq 2$ W/m$^2$, $V_f \approx V_{Df}$ in the range 1.6–2.0 eV and determines the magnitude of the true diffusion photo-emf. Back low- and strong-absorption illuminations photogenerate holes that diffuse to the free PTP surface and decrease $V_{df}$. These illuminations change the polarity of $V_b$. The spectra of the density of short-circuit current $J_{sc}$ of the sandwich cells $SnO_2$ PTPAg and $V_b$ correlate. Under the modulated illumination of these cells through the $SnO_2$ and Ag-electrodes, the drift flow of nonequilibrium holes in the electric field, conditioned by the antiblocking bending of the PTP bands near these electrodes, prevails.

1. Introduction

The discovery of the comparatively low dark resistance $\rho(300\ K) = 100$ Ohm·m in thioderivatives of tetracene [1–3] has stimulated the synthesis of thioderivatives of pentacene [4]. Tetracene and pentacene belong to the homologous series of rectilinear acenes, whose molecules contain four or five benzene rings connected in series, respectively. Moreover, adjacent benzene rings have two common carbon atoms C. The investigation of electrophysical properties of these compounds is urgent from the viewpoint of the search for new photosensitive materials.

Pentacene forms thioderivatives with various numbers of sulfur atoms S. Photovoltaic and optical properties of their mixture, PTP, were investigated in [5]. The most probable components of PTP are tetrathiopentacene (TTP) and hexathiopentacene (HTP). The nature of their long-wave absorption bands was established in [6]. Moreover, field-controlled transistors were fabricated [7], and the photovoltaic properties of thin-film HTP-based heterostructures were investigated [8]. Possible configurations of molecules of pentacene thioderivatives were considered, and it was established that the position of the maximum of long-wave absorption bands of these molecules was described by a linear function of the number of valence electrons of S atoms that took part in the conjugation with the π-system of the pentacene frame of a PTP molecule. The analysis of the spectra of the photocurrent $I_{ph}$ and the capacitor photo-emf (CPE) in the region of the first electron transitions of PTP demonstrated that the photoconductivity was of the p-type and conditioned by dissociation of excitons at electron capture centers [9].

The given paper deals with the investigation of the influence of a permanent illumination on CFE for the purpose to attain a deeper understanding of the nature of the photovoltaic effect in PTP films.

## 2. Experimental Technique

PTP films were obtained by means of thermal deposition in 0.6-mPa vacuum onto quartz substrates at room temperature. Their thickness was measured by an MII-4 interference microscope.

The cells were illuminated through an SPM-2 monochromator by a halogen incandescent lamp or xenon lamp with a power of 100 or 120 W, respectively. The halogen lamp was used to obtain the CPE spectra in the range 1.40–2.48 eV, while the xenon one – in the range 2.48–3.50 eV. The illumination with constant intensity was realized with the help of a DRSh-50 mercury lamp directly or through light filters that transmitted quasimonochromatic spectral domains. The power of permanent or modulated light falling on the measuring cell was measured by means of an RTN-20 thermopile and a pyroreceiver with sensitivities of 0.83 and 250 V/W, respectively.

The CPE was measured in a capacitor cell, in which PTP films deposited onto quartz substrates with a conducting SnO2 layer were separated from an analogous substrate by a transparent dielectric teflon film of 10 μm in thickness. The capacity of the cell amounted to 200 pF. The frequency of the modulated light was equal to 72 Hz.

As a light modulator, we used an aluminum disk with symmetric sector cuttings, whose rotation was accompanied with the formation of rectangular pulses with a duration of 7 ms and a leading-edge time of 20 μs.

The signal was registered with the help of a phase- sensitive amplifier UPI-1 with a digital voltmeter or recorder at the output. The input resistance of the setup was equal to 20 MOhm. A reference signal was applied to the phase detector simultaneously with the amplified CPE signal, which allowed one to monitor the change of its sign.

Sandwich sells MePTPMe and SnO2 PTPMe, where Me=Ag or Al were obtained by successive thermal deposition in 0.6-mPa vacuum onto quartz substrates of the lower electrode, PTP layer, and upper electrode. The construction of these cells allowed one to control the magnitude of resistances of the electrodes. The cells were connected to the input of a UPI-1 or U5-9 amplifier.

The absorption spectra of the PTP films were obtained with the help of a double-beam spectrophotometer "Hitachi" at the spectral width of the gap of 2 nm and room temperature.

The CPE spectra of the capacitor cells and Jsc of the sandwich cells were recalculated to the same power of incident light.

3. Obtained Results and Their Discussion

Figure 1 presents the absorption and CPE spectra of PTP films with the thicknesses of 560 and 1750 nm, respectively. The illumination through the SnO2 - electrode and the teflon film will be considered frontal, while that through the opposite similar electrode that directly contacts with the PTP film – back. From the left, the ordinate axis indicates the magnitude of the potential of the frontal electrode, from the right – the absorption coefficient.

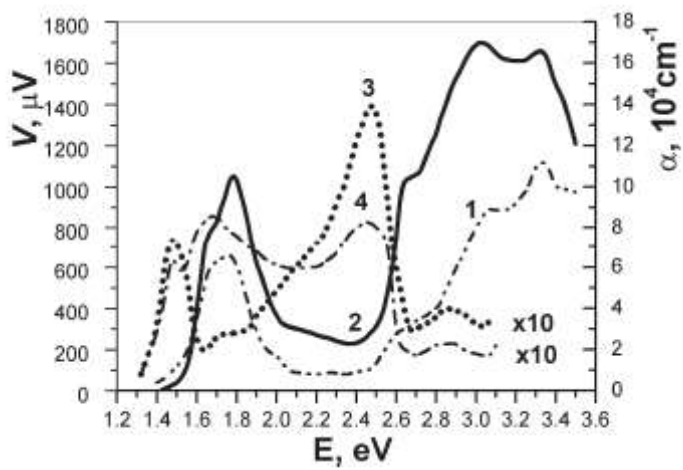

Fig. 1. Absorption spectrum of 0.56-μm PTP film (1), spectra of frontal (2) and back (3) CPE in 13-mPa vacuum and back CPE in air (4) of 1.75-μm PTP film

In 13-mPa vacuum, the frontal CPE spectrum (curve 2) correlates with the absorption spectrum (curve 1) in the region 1.45–3.50 eV. In this case, the free surface of PTP that contacts with the teflon film is charged negatively, while the frontal electrode is charged positively due to electrostatic induction. Back illumination doesn't change the sign of CPE. Its spectrum (curve 3) anticorrelates with the absorption spectrum except for the weak band at 1.77 eV. The absolute value of the back CPE is lower than that of the frontal one by an order of magnitude. The position of its long-wave 1.48-eV band is close to that of the photocurrent band Iph , that was registered in the case of permanent illumination of a surface cell with a PTP layer of 560 nm in thickness [9] (not presented in Fig.1). The spectrum of Iph correlates with the back CPE spectrum in the range 1.45–2.40 eV.

After the filling of air in the measuring cell, the peak intensity of the 1.77-eV band of the frontal CPE decreases by 9% without disturbing its correlation with the absorption spectrum. In this case, the peak intensity of the 1.48-eV band of the back CPE decreases by 13%, whereas that of the band in the region 1.60–1.80 eV increases by a factor of three (curve 4).

The CPE spectra for PTP films with various thicknesses d were measured. In the range of thicknesses 560–1750 nm, the absolute value of the back CPE doesn't change, while that of the frontal CPE slightly decreases with decrease of d. If d < 560 nm, one observes a more abrupt decrease of the frontal CPE against the background of constant back CPE. For d = 190 nm, the frontal CPE is lower than that for films with d ≥ 560 nm by an order of magnitude and comparable with the back CPE in absolute value. The CPE magnitudes were compared by their peak values in the spectral range 1.40–2.40 eV.

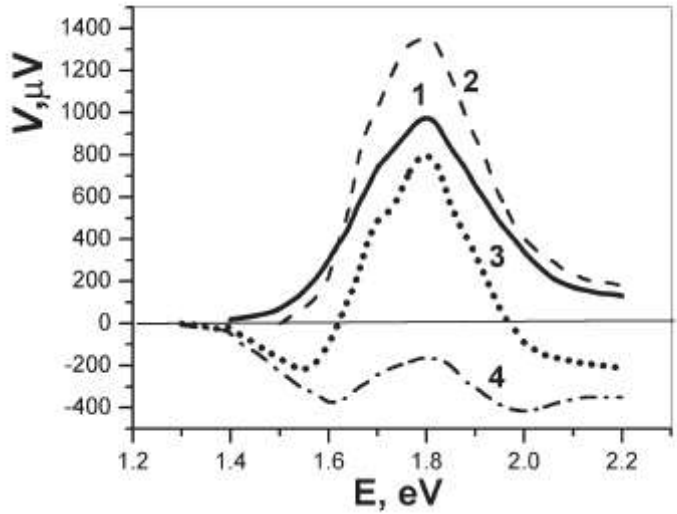

Fig. 2. Spectra of the frontal CPE of a 1.54-μm PTP film in air without illumination (1) and with low- (2) and strong-absorption (3) frontal illuminations with the energies of quanta of 1.54 and 2.64 eV, respectively. The difference of curves 3 and 1 is described by curve 4

In order to determine the regions of strong and weak absorption of PTP films, the Bouguer–Lambert law was used [10]

$$I = I_0 \exp(-\alpha d), \qquad (1)$$

where $I_0$ and $I$ denote the intensities of light that falls on the surface of a PTP film and passes through it, respectively; $\alpha$ is the absorption coefficient; $d$ is the thickness of the film.

The transmission coefficient of PTP films $T = I/I_0$, whereas the relative number of absorbed photons (absorption factor) in the case where their reflection from the surface of the film and scattering in its volume can be neglected $A = 1 - T$. The boundary coefficient $\alpha_0$ that separates the regions of strong and weak absorption was determined with regard for (1) from the condition $A/T = 10$. It was established that $\alpha_0 = 2.398 d^{-1}$. For a PTP film of 1540 nm in thickness, $\alpha_0 = 15600$ cm$^{-1}$. If $\alpha < \alpha_0$, the PTP film absorbs weakly. For $\alpha \geq \alpha_0$, PTP absorption is strong.

The influence of permanent illumination was investigated for films with thicknesses of 1540 and 1750 nm in the region of the first electron transition of PTP. In Fig. 2, curve 1 corresponds to the frontal CPE of the 1540-nm film without illumination. The low-absorption monochromatic 1.54-eV frontal illumination (curve 2) decreases this CPE in the spectral range 1.45–1.60 eV and increases it for quantum energies $1.60 \leq h\nu \leq 2.20$ eV. The strong-absorption 2.67-eV frontal illumination decreases the frontal CPE and, in the region of weak PTP absorption, one observes the change of its sign, that is, the free surface of the film is charged positively, while the frontal electrode is charged negatively (curve 3). Curve 4 is obtained by subtracting curve 1 from curve 3. It reflects the contribution of the frontal strong- absorption illumination to the frontal CPE. Similar changes were observed in the case of strong-absorption illumination with the energy of quanta of 1.73 eV, whose magnitude was close to the energy of the first singlet transition of PTP. The maximum of the negative CPE (by the sign of charge on the frontal electrode) is located at 1.55 eV (curve 4) and is close to that for Iph [9].

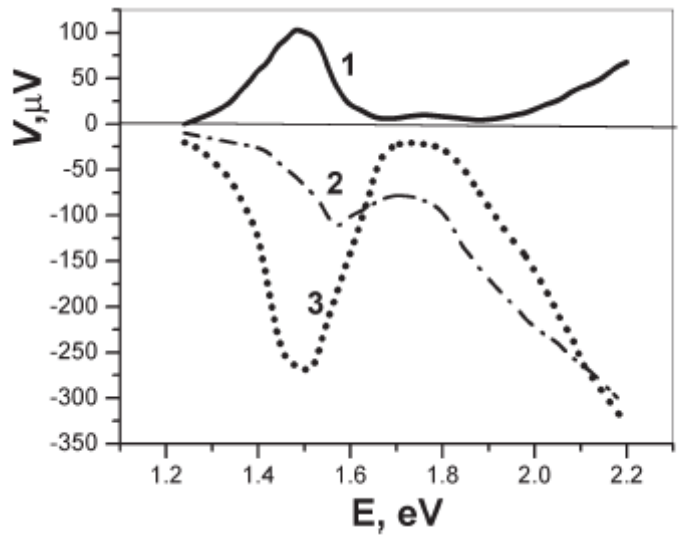

Fig. 3. The same as in Fig. 2 for back CPE

Unfiltered light of a mercury lamp acts as strong- absorption monochromatic illumination. In contrast to analogous 1.54-eV illumination, low-absorption frontal illumination with the energy of quanta of 2.15 eV changes the polarity of the frontal CPE in the region of weak PTP absorption. In the region of strong PTP absorption, all frontal low-absorption monochromatic illuminations act in the same way, that is, they increase the frontal CPE.

Back illuminations (both strong- and low- absorption) increase the frontal CPE without disturbing its correlation with the PTP absorption spectrum. In the case of low-absorption back illumination, the increase of CPE is higher as compared with strong-absorption back illumination, their intensities being the same.

The spectrum of the back CPE without illumination for the PTP film of 1540 nm in thickness in the range 1.20–2.20 eV is presented in Fig. 3 (curve 1). Low- and strong-absorption back illuminations at 1.54 and 2.67 eV, respectively, change the sign of this CPE in the whole spectral region indicated above. The comparison with curve 1 demonstrates that the 1.54-eV illumination (curve 2) hypsochromatically (to the short-wave side) shifts the long-wave maximum of the negative CPE by 0.1 eV. In this case, its peak intensity doesn't change. The back 2.67-eV illumination (curve 3) doesn't change the position of the maximum and increases the peak value of CPE by a factor of 2.8.

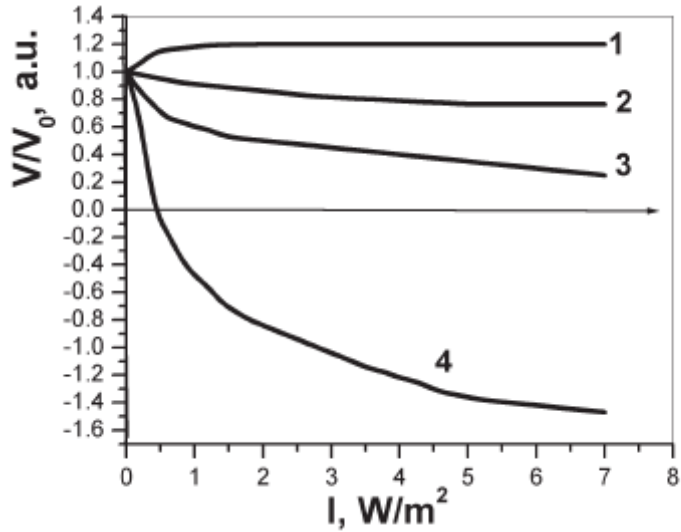

Fig. 4. Dependences of the frontal CPE normalized to the initial value on the intensity of low-absorption frontal 1.54-eV illumination (1, 3) and strong-absorption 1.73-eV illumination (2, 4). Curves 1 and 2 correspond to strong PTP absorption ($E_f$ = 1.77 eV), 3 and 4 – weak PTP absorption ($E_f$ = 1.51 eV)

In all the cases considered above, the intensity of illuminations exceeded that for modulated light by an order of magnitude. Figure 4 presents the dependences of the frontal CPE normalized to the initial value at a specified energy of photons on the intensity of illumination $I_{pi}$. For modulated monochromatic light with the energies of quanta of 1.51 and 1.77 eV, the intensity amounts to 1.32 and 0.94 W/m$^2$, respectively. In the region of strong PTP absorption (1.77 eV), CPE nonlinearly increases as the intensity of low-absorption frontal 1.54-eV illumination rises up to the value $I_{pi}$ = 2 W/m$^2$ (curve 1). A further increase of $I_{pi}$ doesn't change the value of CPE, and curve 1 passes to a horizontal straight line. The strong-absorption frontal 1.73-eV illumination (curve 2) gradually decreases the frontal CPE. In this case, the transition to the horizontal level takes place at $I_{pi}$ ≥6 W/m$^2$. In the region of weak PTP absorption (1.51 eV), a nonlinear decrease of the frontal CPE under the action of low-absorption frontal illumination takes place at $I_{pi}$ ≤2 W/m$^2$ (curve 3). At $I_{pi}$ > 2 W/m$^2$, one observes its rectilinear drop without transition to the horizontal level. One can assume that, in this case, the change of a sign of the frontal CPE is possible at high $I_{pi}$, as it was observed for low-absorption frontal 2.15-eV illumination. The largest changes of CPE take place in the case of weak PTP absorption (1.51 eV) and strong-absorption frontal 1.73-eV illumination

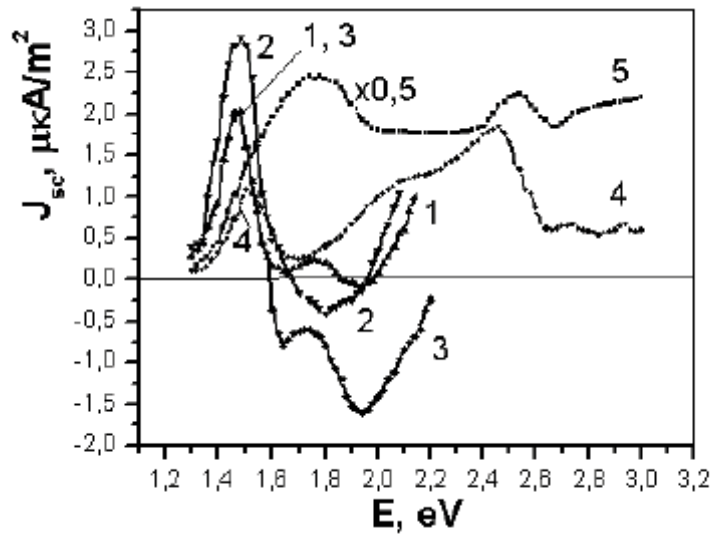

Fig. 5. Spectra of the density of short-circuit current Jsc of the sandwich SnO2 PTPAg cell under modulated lighting through the SnO2 -electrode without illumination (1) and with strong- (2) and low- absorption (3) 2.67-eV and 1.54-eV illuminations through this electrode; AgPTPAl cells without illumination under modulated lighting through the Ag-electrode (4) and Al-electrode (5). The thickness of the PTP layer in all the cells amounted to 1.54 μm

(curve 4). In this case, it nonlinearly decreases with increase of Ipi , changes the sign, and passes to the horizontal level at Ipi1 = 0.4 and Ipi2 ≥7.0 W/m$^2$ , respectively. All nonlinear regions of CPE variation turn into straight lines, if we plot the logarithm of the illumination intensity on the abscissa axis, that is, CPE in these regions is proportional to the logarithm of the illumination intensity.

The spectra of Jsc of the sandwich cells AgPTPAl and SnO$_2$ PTPAg with a PTP layer of 1540 nm in thickness are given in Fig. 5. Illumination through the upper electrodes (Al or Ag) will be considered frontal, whereas that through the lower electrodes (Ag or SnO$_2$ ) – back. In the case of AgPTPAl, the position of the longwave maximum of frontal Jsc (curve 5) coincides with that for the absorption and frontal CPE spectra (Fig. 1, curves 1 and 2, respectively). In this case, the spectra of back Jsc (curve 4) and the back CPE (Fig. 1, curve 3) correlate in the spectral range 1.30–3.4 eV.

For the SnO$_2$ PTPAg cell, the spectrum of back Jsc (curve 1) correlates with the back CPE spectrum (Fig. 3, curve 1). The strong-absorption 2.67-eV back illumination increases the peak intensity of the long- wave band of back Jsc (curve 2). Moreover, in the spectral range 1.68–1.76 eV, back Jsc changes a direction. The low-absorption back 1.54-eV illumination doesn't influence the peak intensity of the long- wave band of back Jsc (curve 3) and changes its direction in the spectral range 1.58–2.26 eV. For the both back illuminations, the change of the direction of Jsc occurs in the region of strong PTP absorption.

The photo-emf V (potential difference across the opposite surfaces of a semiconductor layer) appears under the conditions of illumination of one of the surfaces due to the diffusion-drift motion of free nonequilibrium charge carriers. At the beginning of illumination, one of the components of the flow of nonequilibrium charge carriers (diffusion or drift one) prevails depending on specific conditions. The stationary photo-emf Vst is established when the resulting flow of charge carriers through the semiconductor is equal to zero. If we put electrodes on surfaces of a semiconductor and connect a device to them, the potential difference measured by the latter won't be equal to the value of Vst, as it is influenced by diffusion and drift of charge carriers through the electrodes.

The Bergmann method [11, 12] allows one to avoid the influence of electrodes. According to it, a comparatively thick crystal or a pressed tablet of semiconductor is coated with a thin layer of transparent isolator (glass, mica) from both sides and placed between metal electrodes, one of them being semitransparent. The obtained capacitor cell is illuminated with modulated light through the semitransparent electrode. In this case, the alternating voltage was induced across the electrodes, which, after amplification, is registered by a device. The absolute value of the instantaneous voltage is numerically equal to V. An approximate theory of the Bergmann method is presented in [13].

In our paper, a modified capacitor cell is used, in which one surface of the PTP film is free, while the other contacts with the $SnO_2$-electrode.

The volt-ampere characteristics (VAC) of the dark resistance of sandwich cells SnO2 PTPAg have demonstrated that the SnO2 - and Ag-electrodes are quasiohmic for PTP.

In the modified cell, the capacitor disks were presented by the free surface of PTP and the frontal SnO2 -electrode separated with a transparent teflon film. In [9], it is established that the photoconductivity of PTP is of the p-type. Photogenerated holes diffuse in the bulk of the film from the illuminated free surface of PTP to the back SnO2 -electrode, near which they are accelerated due to the electric field created by the antiblocking bending of PTP energy bands and recombine in the electrode with electrons of the external region of the circuit. In this case, the free surface is charged negatively, while on the frontal electrode, positive charge is induced.

The voltage drop at the input impedance of the device doesn't equal the value of V. The electric circuit of the modified cell and the device represents a series connection of the active resistance of the PTP layer Rc = 155 MOhm, reactance of the cell Xc = 11 MOhm, and the active input resistance of the device Rdev = 20 MOhm and its capacitive resistance Xdev = 44 MOhm connected in parallel. The calculation of this circuit gives the value of the instantaneous voltage Udev at the input impedance of the device:

$$U_{dev} = 0.1V (1 - 0.3i), \qquad (2)$$

where V is the instantaneous value of the photo-emf, I is the imaginary unity.

Analyzing formula (2), one can see that Udev is a complex-valued quantity, in which the active component prevails. The absolute value |Udev| = 0.1 V, i.e. |Udev| << V. This means that the photoeffect in the modified cell is measured in the mode of short-circuit current.

CPE(V) is considered as an algebraic sum of the Dember volume photo-emf (PE) (VD), surface-barrier PE (Vbarr), and surface PE (Vcapt). Vcapt is conditioned by the capture of nonequilibrium charge carriers by surface states of the donor or acceptor type, whereas Vbarr is caused by the separation of free charge carriers by the field of the space-charge region (SCR) [14]. The signs of the quantities Vcapt and Vbarr are determined, respectively, by the type of captured charge carriers and the direction of the bending of energy bands close to the semiconductor surface under equilibrium conditions.

The capture of free charge carriers with surface states influences the magnitude of the near-surface bending of the bands. That's why Vbarr and Vcapt are inseparably linked with each other. In what follows, by Vd, we'll imply the algebraic sum of Vbarr and Vcapt.

The quantity VD is determined by the formula [15]

$$V_D = kT/e \cdot (\mu_n - \mu_p)/(\mu_n + \mu_p) \cdot \ln \sigma_1/\sigma_2, \quad (3)$$

where T denotes the absolute temperature; $e = 1.6 \times 10^{-19}$ C is the elementary charge; $\mu_n$ and $\mu_p$ stand for mobilities of electrons and holes, respectively; $\sigma_1$ and $\sigma_2$ are specific electroconductivities of the semiconductor close to the illuminated and non-illuminated electrodes, respectively.

The analysis of formula (3) shows that $V_D$ appears in the case of the bipolar diffusion of nonequilibrium charge carriers due to a difference of mobilities. If $\mu_p > \mu_n$, the sign of the quantity $V_D$ on the illuminated surface is negative, and it is changed to the positive one if $\mu_p < \mu_n$.

The specific electroconductivity is determined by the formula

$$\sigma = (\mu_n n + \mu_p p)e, \qquad (4)$$

where n and p are concentrations of nonequilibrium electrons and holes, respectively.

In the case of monopolar diffusion, for example, hole one, $\mu_n \ll \mu_p$ and $n\mu_n \ll p\mu_p$. In this case, Eq. (3) is simplified. With regard for (4), we obtain

$$V_D = kT/e \cdot \ln p_1/p_2, \qquad (5)$$

where $p_1$ and $p_2$ are the concentrations of photogenerated holes near the illuminated and non-illuminated surfaces of the semiconductor, respectively.

In the case of illumination of the semiconductor layer with modulated light, the device registers only the variable component V caused by nonequilibrium charge carriers. Let the concentration of nonequilibrium holes change by the exponential law,

$$p = p_0 \exp(-\alpha x), \qquad (6)$$

where $p_0$ is a constant numerically equal to the concentration of nonequilibrium holes close to the illuminated surface of the semiconductor; x is the distance from the illuminated surface to the given point of the semiconductor volume.

At $x_1 = 0$, $p_1 = p_0$ and, at $x_2 = d$ (d is the thickness of the semiconductor layer), $p_2 = p_0 \exp(-\alpha d)$. The substitution of the expressions for $p_1$ and $p_2$ into Eq. (5) simplifies the expression for $V_D$:

$$V_D = - kT/e \cdot \alpha d \qquad (7)$$

The analysis of Eq.(7) shows that, at constant T and d, $V_D \sim \alpha$, i.e. the spectrum of the quantity $V_D$ and the absorption one correlate.

The decrease of the magnitude of the resulting frontal PE $V_f$ following the reduction of the film thickness d as well as correlation of its spectrum with the PTP absorption spectrum (Fig. 1, curves 2 and 1, respectively) according to Eq. (7) indicate that the main contribution to $V_f$ is made by the frontal diffusion component $V_{Df}$. The change of a sign of $V_f$ and the decrease of its value under the action of frontal strong- absorption 1.73-eV and 2.67-eV illuminations (Fig. 2, curve 3; Fig. 4, curves 2 and 4) testify to the fact that $V_{Df}$ is superposed with PE of the opposite sign (Fig. 2, curve 4). The maximum at 1.58 eV in its spectrum is formed due to the surface recombination [14], and its position is close to that in the spectrum of $I_{ph}$ for PTP surface cells [9]. Based on this fact, one can consider that this PE is $V_{df}$.

The depth of light penetration in the bulk of the PTP film $x = 2.398/\alpha$. This quantity for low- and strong- absorption light amounts to: $x_1 = 358$ and $x_2 = 1500$ nm, respectively, i.e. at the thickness of the film $d = 1540$ nm, $x_1$ ï d and $x_2 \approx d$. Taking this fact into account, one can put down: $V_f = V_{Df} - V_{df}$, where $V_{Df}$ and $V_{df}$ are the diffusion and drift components of $V_f$, respectively.

In the spectral range 1.40–1.62 eV (weak PTP absorption), the frontal low-absorption 1.54-eV and 2.15-eV illuminations increase the antiblocking bending of PTP bands (Fig. 2, curve 2; Fig. 4, curve 3). This effect is weaker as compared with that for strong-absorption illuminations. In the case of strong PTP absorption (1.62–2.10 eV region), low-absorption illuminations increase $V_f$ (Fig. 2, curve 2; Fig. 4, curve 1). With increase of their intensity, the spectrum of $V_f$ in the region of strong PTP absorption approaches the spectrum of true $V_{Df}$. At $I_{pi} \geq 2$ W/m2 (Fig. 4, curve 1), this effect saturates and $V_f \approx V_{Df}$, while $V_{df} \approx 0$ (the case of straight bands).

The question about the opposite action of the low- absorption frontal 1.54-eV and 2.67-eV illuminations in the regions of weak and strong PTP absorption remains open.

The increase of Vf under the action of the back low- and strong-absorption 1.54-eV and 2.67-eV illuminations can be explained by a decrease of the antiblocking bending of the bands close to the free PTP surface due to the recombination of holes photogenerated by illuminations with electrons captured with surface states. These holes are photogenerated close to the back electrode and in the bulk of the film and diffuse to the free PTP surface. At the same intensities, the action of low-absorption illumination is stronger, because, in this case, more holes appear in SCR.

The resulting back CPE Vb has all attributes of surface-barrier Vdb . The position of its maximum at 1.48 eV (Fig. 3, curve 1) is close to that for Iph [9]. Vb = Vdb − VDb , where Vdb and VDb are the back barrier and diffusion PE, respectively. The quantities VDf and VDb have opposite signs. One can assume that their absolute values are close as the conditions of illumination are the same in both cases. As Vdb > VDb , the antiblocking bending of the bands near the back $SnO_2$ -electrode is larger than that near the free PTP surface.

Back low- and strong-absorption illuminations decrease the antiblocking bending of the bands near the back electrode and, respectively, the value of Vdb . In this case, VDb is not practically changed and predominates over Vdb , which causes the change of a sign of the resulting Vb . Under the action of the strong- absorption illumination, this change is larger, which is confirmed by a more abrupt spectral dependence of Vb and its higher peak value as compared with those for the low-absorption illumination (Fig. 3, curves 3 and 2, respectively). An abrupt decrease of Vb in the region of strong PTP absorption can be caused by the surface recombination [14]. After turning on the back illuminations, holes photogenerated by modulated light diffuse mainly to the free PTP surface. The device measures Vbst , which is established in the case of positive charge on the free PTP surface and negative charge on the frontal electrode.

The calculation of the electric circuit of the sandwich cell and the device has shown that the instantaneous value of photocurrent is determined by the formula

Iph = Isc (0.90 + 0.04i),        (8)

where Isc = V /Rc is the absolute value of the short- circuit photocurrent.

According to (8), the photocurrent represents a complex-valued quantity, in which the active component prevails.

The absolute value of Iph = 0.90 Isc , i.e. the measurements were performed in the mode of short-circuit current.

The spectral dependences of Jsc of the sandwich cells SnO2 PTPAg and Ag PTPAl illuminated through the $SnO_2$ - and Ag-electrodes (Fig. 5, curves 1 and 4, respectively) are similar to that for Vb of the capacitor cell (Fig. 1, curve 3). It means that the main contribution to the photocurrent is made by the drift flow of nonequilibrium holes in the electric field created by the antiblocking bending of the PTP bands close to the SnO2 - and Ag-electrodes.

The spectral dependence of the quantity Jsc of the sandwich cell AgPTPAl illuminated through the Al-electrode (Fig. 5, curve 5) correlates with the PTP absorption spectrum (Fig. 1, curve 1). One can assume that, near the Al-electrode, the antiblocking bending of the bands is formed for holes, and the directions of the diffusion and drift currents are same, or the surface recombination near this electrode in the case of its quasiohmicity is absent. A more detailed study of the nature of PE near the Al-electrode will represent the subject of further investigations. The difference between the peak values of Jsc is caused by different transmissions of the electrodes.

The illumination of the sandwich cells was performed only through the $SnO_2$ -electrode. The strong- and low- absorption illuminations with the energies of quanta of 2.67 and 1.54 eV (Fig. 5, curve 2 and 3, respectively) change the direction of Jsc in the region of strong PTP absorption. The effect of a change of Jsc is stronger in the case of low-absorption illumination. These data testify to the fact that, under the action of illuminations, the drift direction of Jsc conserves in the spectral range 1.4–1.6 eV (weak PTP absorption) and changes to the diffusion direction in the range 1.6–2.0 eV (strong PTP absorption).

The Vb spectrum was recorded in vacuum (Fig. 1, curve 3). After the filling of air in the cell, a less gap in the CPE spectrum at 1.54 eV (Fig. 1, curve 4) testifies to a decrease of the rate of surface recombination. After keeping the cell in air in darkness, the Vb spectrum changes and can be described by a curve close to that of the initial Vb spectrum measured in vacuum (Fig. 1, curve 3). This fact testifies to changes in PTP films, due to which the energy level distribution of electrons close to the surface of the PTP film in air becomes the same as in vacuum in the course of time.

The influence of the electric field on the magnitude of CPE for capacitor cells will be the subject of further investigations.

4. Conclusions

Based on the investigations performed in the given paper, one can make the following conclusions:

The spectrum of the resulting frontal photo-emf $V_f$ correlates with the PTP absorption spectrum. It is formed by the spectra of the diffusion photo-emf $V_{Df}$ and the drift one $V_{df}$ of opposite signs. In this case, the free surface of the PTP film is charged negatively, whereas, on the frontal $SnO_2$-electrode, a positive charge is induced. The strong-absorption 1.73-eV and 2.67-eV frontal illuminations increase the magnitude of the antiblocking near-surface bending of the bands for holes and, respectively, $V_{df}$, which results in the decrease of $V_f$ in the spectral range 1.4–2.2 eV and the change of its polarity in the range 1.4–1.6 eV (weak PTP absorption). The frontal low-absorption illuminations with the energy of quanta of 1.54 eV and 2.15 eV increase $V_f$ in the range 1.6–2.0 eV due to a decrease of the antiblocking bending of the bands close to the free PTP surface. If the intensities of these illuminations $I_{pi} \geq 2$ W/m$^2$, then $V_f \approx V_{Df}$, i.e. it determines the magnitude of the true diffusion component. Back low- and strong-absorption illuminations photogenerate holes that diffuse to the free surface of PTP and decrease $V_{df}$.

The spectrum of back $V_b$ anticorrelates with the PTP absorption spectrum and correlates with the spectrum of $I_{ph}$. It is formed by the drift photo-emf $V_{db}$ and the diffusion one $V_{Db}$ of opposite polarities. By polarity, $V_b$ coincides with $V_f$ and is lower than the former by an order of magnitude. Moreover, $V_{db} > V_{Db}$. Strong- and low-absorption back illuminations change the sign of $V_b$ in the spectral range 1.4–2.2 eV.

The spectra $J_{sc}$ of the sandwich cells SnO2 PTPAg are similar to those for $V_b$ of the capacitor cells. In the case of illumination of the sandwich cells through the SnO2 - and Ag-electrodes, the drift flow of nonequilibrium holes in the near-electrode electric fields prevails. In the spectral range 1.6–2.2 eV, low- and strong-absorption illuminations change the direction of $J_{sc}$.

ФОТОЕЛЕКТРИЧНИЙ ЕФЕКТ У ПЛІВКАХ ПОЛІТІОПЕНТАЦЕНУ І ВПЛИВ НА НЬОГО ПОСТІЙНОЇ ПІДСВІТКИ

М.П. Горішний

Р е з ю м е

Досліджено внутрішній фотоефект у плівках політіопентацену (ПТП). Виявлено, що у конденсаторних комірках спектральна залежність фронтальної фото-ерс $V_f$ корелює, а тильної фото-ерс $V_b$ антикорелює із спектром поглинання ПТП. Полярності $V_f$ і $V_b$ збігаються. При цьому заряд вільної поверхні ПТП негативний, а фронтального $SnO_2$ -електрода – позитивний. $V_f = V_{Df} - V_{df}$ і $V_b = V_{db} - V_{Db}$, де $V_{Df}$ і $V_{Db}$ – дифузійні, $V_{df}$ і $V_{db}$ – дрейфові фронтальна і тильна складові відповідно. Величини $V_{df}$ і $V_{db}$ зумовлені антизапірними вигинами зон для дірок біля вільної поверхні ПТП і тильного $SnO_2$-електрода відповідно. Сильнопоглинальна монохроматична фронтальна підсвітка 1,73 і 2,67 еВ збільшує модуль $V_{df}$, що зменшує результуючу $V_f$ у спектральному діапазоні 1,4–2,2 еВ і змінює її полярність у діапазоні 1,4–1,6 еВ. Слабопоглинальна фронтальна підсвітка квантами 1,54 і 2,15 еВ зменшує $V_{df}$ у спектральному діапазоні 1,6–2,0 еВ і збільшує її у діапазоні 1,4–1,6 еВ. Якщо інтенсивність цих підсвіток $I_{pi} \geq 2$ Вт/м2, то $V_f \approx V_{Df}$ у діапазоні 1,6–2,0 еВ і визначає величину істинної дифузійної фото-ерс. Тильні слабо- і сильнопоглинальні підсвітки фотогенерують дірки, які дифундують до вільної поверхні ПТП і зменшують $V_{df}$. Ці підсвітки змінюють полярність $V_b$. Спектри густини струму короткого замикання $J_{sc}$ сандвічних комірок $SnO_2$ ПТПAg і величина $V_b$ корелюють. В разі модульованого освітлення цих комірок через $SnO_2$- і Ag- електроди переважає дрейфовий потік нерівноважних дірок в електричному полі, зумовленому антизапірним вигином зон ПТП біля цих електродів.